\documentclass[12pt]{article}

\usepackage[T1]{fontenc}
\usepackage[utf8]{inputenc}
\usepackage{geometry}
\usepackage{amsmath,amssymb}
\usepackage{hyperref}
\hypersetup{
  colorlinks=true,
  linkcolor=blue,
  citecolor=blue,
  urlcolor=blue
}
\usepackage{url}
\usepackage{booktabs}
\usepackage{enumitem}
\usepackage{parskip}
\usepackage[numbers,sort&compress]{natbib}
\title{\textbf{Statistics 101, 201, and 202: Three Shiny Apps for Teaching
Probability Distributions, Inferential Statistics, and Simple Linear
Regression}}

\author{Antoine Soetewey\\[4pt]
\small HEC Li\`ege, ULi\`ege, Rue Louvrex~14, 4000 Li\`ege, Belgium\\
\small \href{https://orcid.org/0000-0001-8159-0804}{ORCID: 0000-0001-8159-0804}\\
\small \href{mailto:antoine.soetewey@uliege.be}{antoine.soetewey@uliege.be}}

\date{30 March 2026}

\begin{document}

\maketitle

\begin{abstract}
Statistics 101, 201, and 202 are three open-source interactive web
applications built with R \citep{R} and Shiny \citep{shiny} to support
the teaching of introductory statistics and probability. The apps help
students carry out common statistical computations---computing
probabilities from standard probability distributions, constructing
confidence intervals, conducting hypothesis tests, and fitting simple
linear regression models---without requiring prior knowledge of R or any
other programming language. Each app provides numerical results, plots
rendered with \texttt{ggplot2} \citep{ggplot2}, and inline mathematical
derivations typeset with MathJax \citep{cervone2012mathjax}, so that
computation and statistical reasoning appear side by side in a single
interface. The suite is organised around a broad pedagogical progression:
Statistics~101 introduces probability distributions and their properties;
Statistics~201 addresses confidence intervals and hypothesis tests; and
Statistics~202 covers the simple linear model. All three apps are freely
accessible online and their source code is released under a CC-BY-4.0
license.
\end{abstract}

\section{Summary}

Statistics 101, 201, and 202 are three open-source interactive web
applications built with R \citep{R} and Shiny \citep{shiny} to support
the teaching of introductory statistics and probability. The apps are
intended to help students carry out common statistical computations
(computing probabilities from standard probability distributions,
constructing confidence intervals, conducting hypothesis tests, and
fitting simple linear regression models) without requiring prior
knowledge of R or any other programming language. Each app provides
numerical results, plots rendered with \texttt{ggplot2} \citep{ggplot2},
and inline mathematical derivations typeset with MathJax
\citep{cervone2012mathjax}, so that computation and statistical reasoning
appear side by side in a single interface.

The suite was developed during the author's doctoral research and
teaching assistantship at UCLouvain and continues to be used as an
invited lecturer at UCLouvain and UNamur in courses on probability,
inferential statistics, and linear regression. The three apps are
organised around a broad pedagogical progression: Statistics~101
introduces probability distributions and their properties; Statistics~201
addresses confidence intervals and hypothesis tests; and Statistics~202
covers the simple linear model. All three apps are freely accessible
online and their source code is released under a CC-BY-4.0 license.

\section{Statement of Need}

Introductory statistics courses in economics, biostatistics, social
sciences, and related disciplines commonly ask students to evaluate
probabilities, construct confidence intervals, perform hypothesis tests,
and fit linear regression models. Working through these computations by
hand can be time-consuming and error-prone, while doing so in R
presupposes a level of programming familiarity that beginners may not yet
have. Printed or online statistical tables can help with probability
computations, but typically cover only a limited number of distributions
and offer no graphical feedback.

Several existing tools partially address this gap. Calculators bundled
with commercial textbooks are often proprietary. General-purpose
environments such as R, Python, or SPSS require students to write or
adapt code. Web-based probability calculators (for instance those offered
by Wolfram Alpha or individual course websites) tend to cover only a
limited number of distributions, and many do not include hypothesis
testing or regression modules, nor do they display the underlying
mathematical expressions alongside their results. Shiny-based tools for
teaching statistics have been explored in the literature
\citep{potter2016web,soetewey2026associationexplorer}, though these tend
to focus on specific topics or workflows rather than covering the full
introductory sequence in single interfaces.

The Statistics~101/201/202 suite is an attempt to complement these
resources in a few ways. It brings together topics from a typical
two-semester introductory statistics sequence (probability distributions,
inference, and regression) within a single, consistent interface. Each
numerical result is shown alongside a dynamically updated plot and the
corresponding mathematical expression, with the aim of supporting
statistical intuition alongside computation. The apps are fully open
source, so other instructors are welcome to adopt, fork, or extend them.
Their continued use across introductory and intermediate statistics
courses at UCLouvain and UNamur suggests they may be useful beyond the
context in which they were originally developed.

\section{Description of the Apps}

\subsection{Statistics 101: Probability Distributions}

Statistics~101
(\url{https://github.com/AntoineSoetewey/statistics-101}) is designed to
help students compute probabilities from 18 probability distributions:
Beta, Binomial, Cauchy, Chi-square, Exponential, Fisher, Gamma,
Geometric (two parameterisations), Hypergeometric, Logistic, Log-Normal,
Negative Binomial (two parameterisations), Normal, Poisson,
Student's~$t$, and Weibull. The app uses a single-page sidebar layout.
Students select a distribution from a drop-down menu, and the sidebar
dynamically renders the appropriate parameter inputs using conditional
panels.

For each distribution, the student specifies the type of probability to
compute:
\begin{itemize}[noitemsep]
  \item Lower tail: $P(X \leq x)$
  \item Upper tail: $P(X > x)$
  \item Interval: $P(a \leq X \leq b)$
\end{itemize}

The main panel displays three outputs:
\begin{enumerate}[noitemsep]
  \item \textbf{Solution:} the probability statement in full mathematical
    notation rendered via MathJax, for example
    $X \sim \mathcal{N}(\mu = 0,\, \sigma^2 = 1)$ and
    $P(X \leq 1) = 0.8413$, computed to four decimal places using R's
    built-in distribution functions (\texttt{pbeta}, \texttt{pbinom},
    \texttt{pnorm}, etc.).

  \item \textbf{Plot:} a density or mass function plot with the relevant
    region shaded, rendered using \texttt{ggplot2} graphics and displayed
    with a loading spinner via \texttt{shinycssloaders}
    \citep{shinycssloaders}.

  \item \textbf{Details:} the probability density function (PDF) or
    probability mass function (PMF) in display-style mathematical
    notation, followed by the expectation $E(X)$, standard deviation
    $SD(X)$, and variance $\mathrm{Var}(X)$, all evaluated numerically
    at the current parameter values.
\end{enumerate}

\subsection{Statistics 201: Confidence Intervals and Hypothesis Tests}

Statistics~201
(\url{https://github.com/AntoineSoetewey/statistics-201}) is intended to
support the teaching of confidence intervals and hypothesis tests across
seven inference settings, selectable from a drop-down menu:
\begin{itemize}[noitemsep]
  \item One mean: $z$-test (population variance $\sigma^2$ known) or
    one-sample $t$-test ($\sigma^2$ unknown)
  \item Two means, independent samples: $z$-test or two-sample $t$-test
    with either equal ($\sigma^2_1 = \sigma^2_2$) or unequal
    ($\sigma^2_1 \neq \sigma^2_2$) variances (Welch's correction)
  \item Two means, paired samples: paired $t$-test or $z$-test on the
    differences $D_i = X_{1i} - X_{2i}$
  \item One proportion: one-sample $z$-test
  \item Two proportions: two-sample $z$-test with an option to use a
    pooled standard error
  \item One variance: $\chi^2$ test, using
    \texttt{EnvStats::varTest()} \citep{EnvStats} for raw data and a
    custom helper function for summary statistics ($n$ and $s^2$)
  \item Two variances: $F$-test, using \texttt{var.test()} from base R
\end{itemize}

For all mean and variance settings, students may provide either raw data
(comma-separated values pasted into a text field) or summary statistics
($n$, $\bar{x}$, $s^2$, etc.) directly. The significance level $\alpha$
is controlled via a slider (range 0.01 to 0.20, default 0.05). The null
hypothesis value $H_0$ and the direction of the alternative hypothesis
(two-sided, greater, or less) are set in the sidebar.

For each procedure, the main panel displays a structured four-step
output:
\begin{enumerate}[noitemsep]
  \item \textbf{Data:} the entered observations (if raw data were
    provided) together with the relevant summary statistics
    ($n$, $\bar{x}$, $s$, etc.), typeset in MathJax.

  \item \textbf{Confidence interval:} the interval formula with all
    quantities substituted numerically, for example
    $\bar{x} \pm \left(t_{\alpha/2,\, n-1} \times s/\sqrt{n}\right) =
    [\ell;\, u]$, in the appropriate one- or two-sided form.

  \item \textbf{Hypothesis test:} a numbered four-step procedure
    consisting of (1)~the statement of $H_0$ and $H_1$, (2)~the test
    statistic formula evaluated numerically, (3)~the critical value(s),
    and (4)~the reject or do-not-reject conclusion.

  \item \textbf{Interpretation:} a plain-language conclusion at the
    chosen significance level, including the $p$-value.
\end{enumerate}

Below the text output, a \texttt{ggplot2} plot shows the relevant
sampling distribution ($t$, $\mathcal{N}(0,1)$, $\chi^2$, or $F$) with
the rejection region shaded and the observed test statistic marked by a
vertical line.

\subsection{Statistics 202: Simple Linear Regression}

Statistics~202
(\url{https://github.com/AntoineSoetewey/statistics-202}) is designed to
help students fit a simple linear regression model
$y = \beta_0 + \beta_1 x + \varepsilon$ to data entered directly in the
browser. The app accepts comma-separated numeric values for the predictor
$x$ and the response $y$, checks that the two vectors are of equal
length and that $x$ contains more than one distinct value. The fitted
model is computed via \texttt{lm()} in R.

The main panel presents the following sequence of outputs:
\begin{enumerate}[noitemsep]
  \item \textbf{Data table:} an interactive \texttt{DT}
    \citep{DT} table of the entered observations, with export buttons
    (copy, CSV, Excel, PDF, print).

  \item \textbf{Summary statistics:} sample means $\bar{x}$, $\bar{y}$,
    and sample size $n$.

  \item \textbf{Step-by-step derivation:} the OLS estimator
    $\hat{\beta}_1$ computed via the formula
    \[
      \hat{\beta}_1 =
        \frac{\bigl(\sum_{i=1}^n x_i y_i\bigr) - n\bar{x}\bar{y}}
             {\sum_{i=1}^n (x_i - \bar{x})^2}
    \]
    followed by $\hat{\beta}_0 = \bar{y} - \hat{\beta}_1 \bar{x}$, both
    rendered in MathJax and evaluated numerically on the student's own
    data \citep{fox2015applied}.

  \item \textbf{R output:} the full \texttt{summary(lm(\ldots))} output,
    including coefficient estimates, standard errors, $t$-statistics,
    $p$-values, and adjusted $R^2$.

  \item \textbf{Interactive regression plot:} a scatter plot with the
    fitted line and an optional confidence band, rendered with
    \texttt{ggplot2} and made interactive via \texttt{plotly}
    \citep{plotly}.

  \item \textbf{Interpretation:} a plain-language reading of
    $\hat{\beta}_0$ and $\hat{\beta}_1$, adapted to the user-supplied
    axis labels and conditioned on whether each coefficient is
    significantly different from zero at the 5\% level.

  \item \textbf{Assumption diagnostics:} a collapsible panel containing
    the six model-checking plots produced by
    \texttt{performance::check\_model()} \citep{performance}, covering
    linearity, homoscedasticity, influential observations and normality
    of residuals.

  \item \textbf{Downloadable report:} an HTML report of the analysis,
    generated from an R~Markdown
    \citep{rmarkdown2018,rmarkdown2020,rmarkdown2025} template, with an
    option to include or hide the underlying R code.
\end{enumerate}

\section{Pedagogical Design}

The three apps share a common design orientation. The interface is kept
relatively minimal so that attention can remain on the statistical
content rather than on navigating the software. No registration,
installation, or command-line interaction is required; students open a
URL and begin working straight away. Results are displayed at a level of
mathematical abstraction close to what is used in lectures, with
notation, formulas, and numerical output appearing together, so that the
apps can serve as a complement to, rather than a replacement for, the
conceptual material covered in class.

The step-by-step derivation in Statistics~202 was motivated by a
difficulty that often arises in introductory regression courses: students
may be able to write down a formula from a slide yet find it hard to
connect it to the numerical output produced by R. Showing both the
symbolic formula and its numerical evaluation side by side, on the
student's own data, is one way of trying to make that connection more
concrete. A related idea underpins the Details panel in Statistics~101,
which displays the PDF or PMF alongside the computed probability, so that
students can see how the shaded area in the plot corresponds to the
analytical expression.

A comparable motivation applies to Statistics~201. Introductory
statistics courses typically cover a large number of inference
procedures, each with its own test statistic formula, degrees of
freedom, and critical value. Students often find it difficult to locate
the relevant formula quickly when working through an exercise,
particularly when the course material is spread across a lengthy syllabus
or a large set of slides. By displaying the appropriate formula for each
procedure alongside its numerical evaluation, Statistics~201 is intended
to serve as a single reference point where students can find, read, and
apply the correct expression without having to search through their
course materials.

The apps were originally developed in response to a practical difficulty
observed during teaching-assistant sessions at UCLouvain, where students
were spending a considerable share of class time on the mechanics of
looking up or computing probabilities and test statistics, leaving
relatively little time for interpretation and discussion. The hope was
that shifting some of that computational work to a transparent,
easy-to-use tool might free up more time for conceptual engagement.

\section{Availability and Reproducibility}

All three apps are hosted on \href{https://www.shinyapps.io/}{shinyapps.io}
and their source code is publicly available on GitHub under a CC-BY-4.0
license:
\begin{itemize}[noitemsep]
  \item Statistics~101:
    \url{https://github.com/AntoineSoetewey/statistics-101}
  \item Statistics~201:
    \url{https://github.com/AntoineSoetewey/statistics-201}
  \item Statistics~202:
    \url{https://github.com/AntoineSoetewey/statistics-202}
\end{itemize}

Each app's GitHub repository includes a README file with instructions on
how to use the app, the dependencies, instructions for local deployment,
and a link to the live version.

\section*{Acknowledgements}

The author thanks Prof.\ Dominique Deprins and Prof.\ Eugen Pircalabelu
for their comments on previous versions of the apps. The author is
grateful to his PhD supervisors, Prof.\ Catherine Legrand and
Prof.\ Michel Denuit, for their support and for allowing him to work on
these apps during his doctoral studies at UCLouvain. Finally, the author
thanks the students at UCLouvain, without whom these apps would not
exist.

\bibliography{paper}

\end{document}